\documentclass[12pt,showkeys,prab,superscriptaddress,floatfix,nofootinbib,longbibliography]{revtex4-2}

\usepackage{graphicx}
\usepackage{color}
\usepackage{amsfonts}
\usepackage{bm}
\usepackage{xspace}

\usepackage[normalem]{ulem}

\usepackage{ORCIDinREVTeX}

\renewcommand {\d} {{\rm d}}

\newcommand {\om} {\omega}

\newcommand{\MBNExplorer} {\textsc{MBN Explorer}\xspace}

\begin{document}

\title{\textcolor{black}{The impact of experimental conditions on
the observation of
channeling and crystalline undulator radiation}}

\author{Maykel M\'arquez-Mijares}
\email[]{mmarquez@instec.cu}
\affiliation{Instituto Superior de Tecnolog\'ias y Ciencias Aplicadas,
University of Havana (InSTEC-UH), Ave. Salvador Allende 1110, Plaza de
la Revoluci\'on, Havana - 10400, Cuba}

\author{Germ\'an Rojas-Lorenzo}
\email[]{grojas@instec.cu}
\affiliation{Instituto Superior de Tecnolog\'ias y Ciencias Aplicadas,
University of Havana (InSTEC-UH), Ave. Salvador Allende 1110, Plaza de
la Revoluci\'on, Havana - 10400, Cuba}

\author{Jes\'us Rubayo-Soneira}
\email[]{jrubayo@gmail.com}
\affiliation{Instituto Superior de Tecnolog\'ias y Ciencias Aplicadas,
University of Havana (InSTEC-UH), Ave. Salvador Allende 1110, Plaza de
la Revoluci\'on, Havana - 10400, Cuba}

\author{Thu Nhi Tran Caliste}
\email[]{thu-nhi.tran-thi@esrf.fr}
\affiliation{European Synchrotron Radiation Facility,
71 Avenue des Martyrs, Grenoble, F-38000,France}

\author{Andrei V. Korol}
\email[]{korol@mbnexplorer.com}
\affiliation{MBN Research Center, Altenh\"{o}ferallee 3, 60438 Frankfurt am Main, Germany}

\author{Andrey V. Solov'yov}
\email[]{solovyov@mbnresearch.com}
\affiliation{MBN Research Center, Altenh\"{o}ferallee 3, 60438 Frankfurt am Main, Germany}

\begin{abstract}
	
In this study, we present a comprehensive quantitative analysis of the
radiation emitted by 855 MeV electrons propagating through an oriented
diamond hetero-crystal.
The crystal consists  of two distinct segments: (i) a straight
single-crystal diamond substrate, and (ii) a diamond layer that is
periodically doped with  boron atoms.
The doping profiles were derived from precise experimental
measurements of boron concentration obtained during the layer’s
fabrication via Microwave Plasma Chemical Vapor Deposition (MPCVD).
Our study systematically investigates the channelling
and the crystalline undulator radiation, accounting for the
different doping profiles in the undulating region.
The simulations were conducted using the advanced \MBNExplorer
software package, which enables detailed modeling of particle
trajectories and radiation emission.
\textcolor{black}{We report on good agreement with experiment and
discuss remaining discrepancies providing possible explanations for
them.}
The results obtained show that the radiation intensity is significantly
affected by a range of factors,
\textcolor{black}{including the angular divergence of the incident
beam, its orientation with respect to the target,}
the direction in which the emitted radiation is detected,
and the choice of the doping profiles.
These findings are important for optimising the design of crystalline undulators
as novel gamma radiation light sources.
\end{abstract}

\maketitle

\section{Introduction}\label{Introduction}

The interaction of high-energy charged particles with crystals is highly
sensitive to the direction of the incoming beam relative to the main
crystallographic axes of the target crystal.
Projectiles that strike a crystal at small angles with respect to its
planes (or axes), can travel long distances within the crystalline
medium.
This particular motion, known as channelling, is due to the collective
action of the electrostatic fields of the lattice atoms \cite{Lindhard}.
The study of the channelling process of ultra-relativistic projectiles
in oriented crystals has become a broad research field
\cite{Uggerhoj:RPM_v77_p1131_2005,BiryukovChesnokovKotovBook,%
ChannelingBook2014,ScandaleTaratin:PhysRep_v815_p1_2019,CLS-book_2022},
with applications including beam steering
\cite{MazzolariEtAl:EPJC_v78_p720_2018}, collimation
\cite{AfoninEtAl:PRL_87_094802_2001}, focusing
\cite{ScandaleEtAl:PRAB_v21_014702_2018}, and extraction
\cite{SytovEtAl:EPJC_v82_187_2022}.

When irradiated by beams of ultra-relativistic electrons and positrons,
oriented crystals of various geometries have the potential to serve as
advanced, versatile crystal-based light sources (CLS).
These structures can generate gamma radiation of high intensity spanning
a broad range of photon energies from MeV to GeV
\cite{CLS-book_2022,SushkoKorolSolovyov:EPJD_v76_166_2022,%
KorolSolovyov:NIMB_v537_p1_2023}.
Recent research \cite{KorolSushkoSolovyov:PRAB_v27_p100703_2024}
has demonstrated that the photon fluxes produced in such systems can
significantly exceed those achieved by the most modern gamma-ray
light sources, such as those based on laser-Compton interactions.
This capability makes oriented crystals a promising alternative for
developing technologies for generating high-energy gamma radiation,
opening up new possibilities for advanced scientific and technological
applications.

One of the proposed schemes for implementing CLS relies on a crystalline
undulator (CU), which consists of an oriented, periodically bent crystal
(PBC), through which a beam of ultra-relativistic charged particles
traverse via channelling motion
\cite{KSG1998,KSG_review2004,ChannelingBook2014}.
The periodic bending results in the crystalline undulator radiation
(CUR) being emitted in addition to the channelling radiation
\cite{ChRad:Kumakhov1976}.
The advantage of PBCs is that the characteristics of the
emitted radiation (its peak intensity and frequency) can be optimised
by adjusting \textcolor{black}{bending amplitude and period}
to the parameters of the incident beam.

PBCs can be fabricated using various techniques, each of which
is suited to the manufacture of a different type of crystal.
In the field of CLS, particular attention has been given to diamond
crystals doped periodically with boron atoms, which can be
produced using microwave plasma chemical vapour deposition
\cite{ConnellEtAl_CVD_2015,ThuNhiTranThi_JApplCryst_v50_p561_2017,%
BackeEtAl:NIMA_v1073_170236_2025}.

Recently, a general formalism was presented that relates the bending
profile of crystal planes to the variation in dopant atom concentration
within  a boron-doped layer \cite{KorolSolovyov:NIMA_v1083_171160_2026}.
It was demonstrated that, in order to maximise both the number of
particles accepted in channelling mode at the crystal entrance and
the collection of CUR, the incident particle beam must be aligned
with the tangent line to the profile, and the detector must be
positioned along the centerline of the periodically bent profile.
These conditions were clearly formulated in Ref.
\cite{KorolSolovyov:NIMA_v1083_171160_2026} and
were applied to simulate the CUR emission from 855 MeV electrons
incident on a quasi-periodically bent diamond crystal, as used in the
experiment \cite{BackeEtAl:NIMA_v1073_170236_2025}.
These conditions were not fully met in the experiment, where the CUR
signal was not detected.

The first experimental observation of the characteristic peak of CUR was
recently reported for an 855 MeV electron beam channelled in a
periodically bent diamond boron-doped crystal oriented along the
(110) \textcolor{black}{plane} \cite{BackeEtAl:arXiv_2504.17988}.
In this context, the main objective of our study is for a
rigorous analysis to be performed on the experimental results.
For doing this, the methodology developed in Ref.
\cite{KorolSolovyov:NIMA_v1083_171160_2026} is applied to carry out
numerical simulations of the electron channeling and photon emission
processes occurring in a crystalline sample used in the experiment.
We analyse the dependence of the spectra of emitted radiation on a
number of angular variables that determine the beam-crystal and the
crystal-radiation geometries.
Some of these angles were specified in the description of the
experimental setup provided in the cited paper whereas some of them were
either not quoted explicitly or quoted without indicating the
uncertainties.
The results obtained show that the radiation intensity is significantly
affected by a range of factors, including the orientation and angular
divergence of the incident beam, the direction in which the radiation is
detected.
\textcolor{black}{We demonstrate good agreement between the results of
numetcal simulations with the data obtained
experimentally.
We provide possible explanations for the remaining discrepancies
between theory and experiment.}

The trajectories of ultra-relativistic electrons in a crystalline
medium were simulated numerically with an atomistic level of accuracy.
This was done within the relativistic classical molecular dynamics
(Rel-MD) framework
\cite{MBN_ChannelingPaper_2013,KorolSushkoSolovyov:EPJD_v75_p107_2021},
using the multi-purpose computer package MBN Explorer
 \cite{MBNExplorer_2012}.
 The implemented algorithms enable the modelling of particle movement
 over macroscopic distances with atomistic accuracy, taking into account
 the interaction between a projectile and all the surrounding atoms.
Particle trajectories are generated by accounting for randomness in the
sampling of projectiles and the displacements of atoms due to thermal
vibrations.
Another phenomenon that affects the dynamics of a projectile particle
is ionising collisions, which lead to a random change in the particle’s
velocity.
As these quantum events occur randomly on the atomic scale in terms of
time and space, they are incorporated into classical framework in
accordance with their probabilities
\cite{SushkoKorolSolovyov:NIMB_v569_165911_2025}.
All simulated trajectories are statistically independent and can be used
to analyse channeling efficiency and characterise emitted radiation
through the calculation of its spectral-angular and spectral
distributions.

Section \ref{Methodology} provides an overview of the methodology
employed in this study, including the target parameters and the
simulation details.
Section \ref{Results} presents a description of the crystalline target
used in the experiment and the numerical results obtained.
Where applicable, the results are compared with the experimental
data collected at the MAinz MIkrotron (MAMI)
facility \cite{BackeEtAl:arXiv_2504.17988}.
The analysis also covers the simulated trajectories and the emitted
radiation as a function of the crystal geometry and the divergence of
the incident beam.
Finally, Section \ref{Conclusion} summarises the main findings and
outlines future prospects.

\section{Methodology}\label{Methodology}

To establish the correspondence between the parameters of the periodic
bending in a crystalline structure and the periodic variation of dopant
atom concentration along a specific crystallographic direction,
we adopt the approach outlined in detail in Ref.
\cite{KorolSolovyov:NIMA_v1083_171160_2026}.
The cited paper presents simulations of electron and positron
channelling, and of the photon emission process, which were performed
under the experimental conditions described in
Ref. \cite{BackeEtAl:NIMA_v1073_170236_2025}

In this contribution, we focus on a hetero-crystal that was used in
a more recent experiment \cite{BackeEtAl:arXiv_2504.17988} with an
855 MeV electron beam.
The hetero-crystal consists of two distinct parts:
\begin{itemize}
 \item[(i)]
a  (100) surface-oriented 194 $\mu$m thick diamond substrate,
 \item[(ii)]
a boron-doped diamond layer, in which the concentration $n_B$ of
boron atoms varies periodically along the [100] crystallographic
direction, (referred to below as the $Z$ direction) with a period
$\lambda_{\rm B}\approx 3.54$ $\mu$m.
The layer accommodates four periods.
\end{itemize}

A change in boron concentration induces a change in the lattice
constant, $a_\perp$, in the $Z$ direction
\textcolor{black}{(see Figure 1
in Ref.~\cite{KorolSolovyov:NIMA_v1083_171160_2026})}.
In contrast, the lattice constant measured within the (100) plane remains at
its equilibrium value of $a_0 = 3.567\,$\AA, which is characteristic
of a single diamond crystal.

Following Refs. \cite{WojewodaEtAl-DiamondRelMat_v17_p1302_2008,BrazhkinEtAl-PRB_v74_140502_2006,BrunetEtAl-DiamondRelMat_v7_p869_1998},
one can assume linear dependence between $a_\perp$ and the boron concentration:
\begin{equation}
a_\perp(Z)=a_0 \Bigl( 1+\kappa n_B(Z) \Bigr)\,.
\label{eq:01}
\end{equation}
The following two values of the coefficient $\kappa$ can be calculated
using the data available in literature:
$\kappa_{\rm B} = 5.38 \times 10^{-25}\,\mathrm{cm}^3$ due to Brazhkin
\textit{et al.} \cite{BrazhkinEtAl-PRB_v74_140502_2006},
and $\kappa_{\rm V} = 8.12 \times 10^{-25}\,\mathrm{cm}^3$ due to Vegard
\cite{Vegard:ZPhysik_v5_p17_1921}.

The change in $a_{\perp}$ leads to a distortion of the crystallographic
planes other than the (100) planes.
In particular, the bending profile of the (110) plane can be related
to $n_B(Z)$ as follows \cite{KorolSolovyov:NIMA_v1083_171160_2026}:
\begin{eqnarray}
y(z) = - {\kappa \over\sqrt{2}}  \int_{0}^{z/\sqrt{2}}n_B(Z')dZ'
\label{eq:profile}
\end{eqnarray}
where the $y$ axis aligns with the [110] axial direction and the $z$
axis lies within the (110) plane in the substrate.

\begin{figure*}[h]
\centering
\includegraphics[clip, width = 0.75\textwidth]{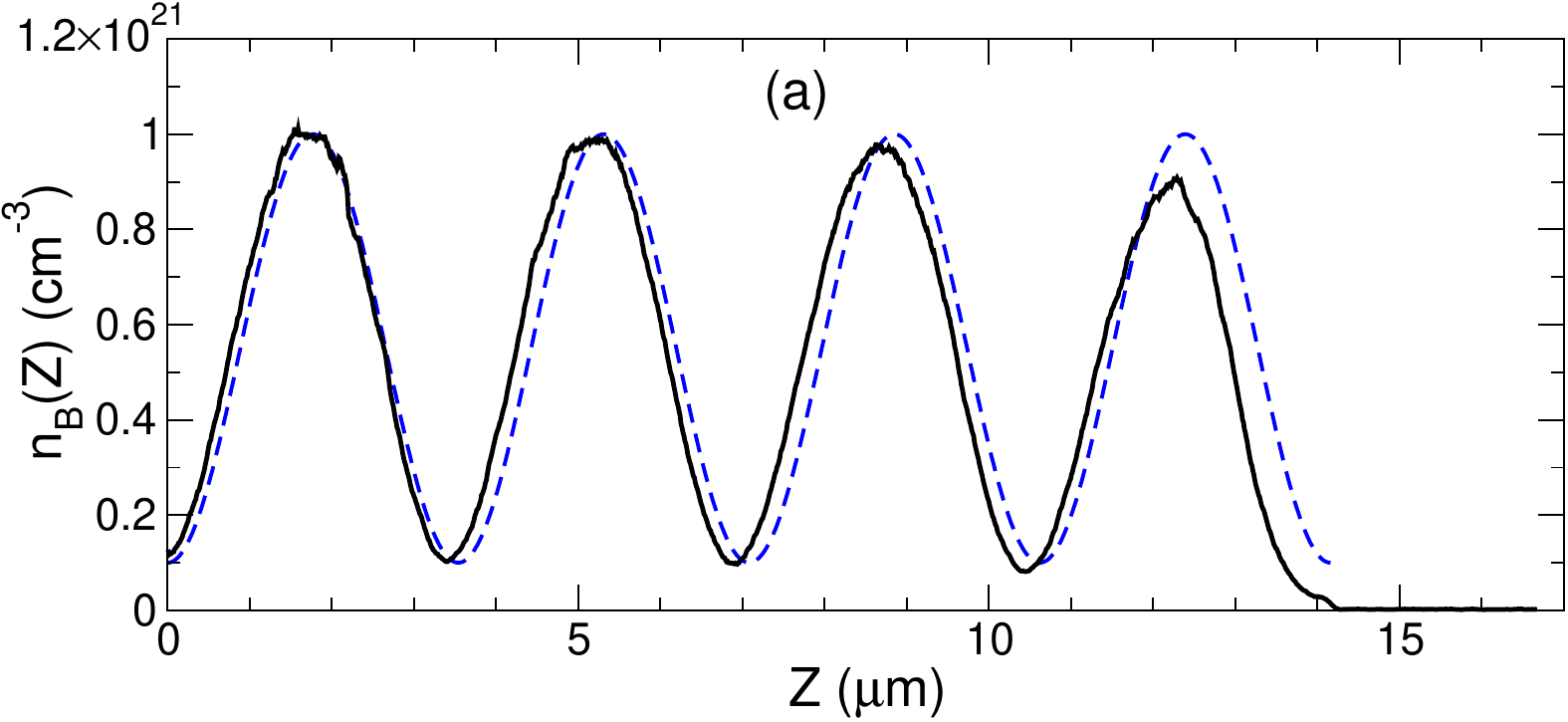}\\
\vspace*{0.25cm}
\includegraphics[clip, width = 0.75\textwidth]{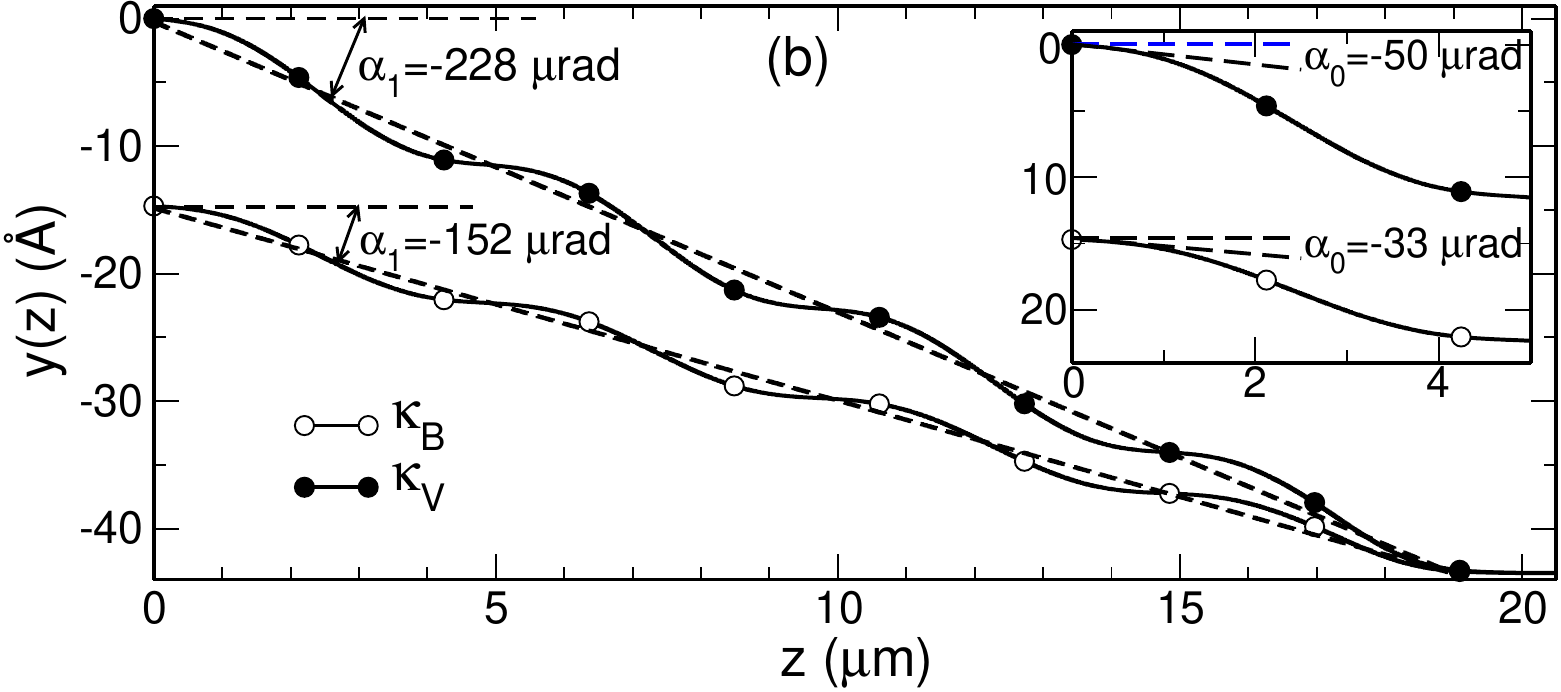}
\vspace*{0.5cm}
\includegraphics[clip, width = 0.75\textwidth]{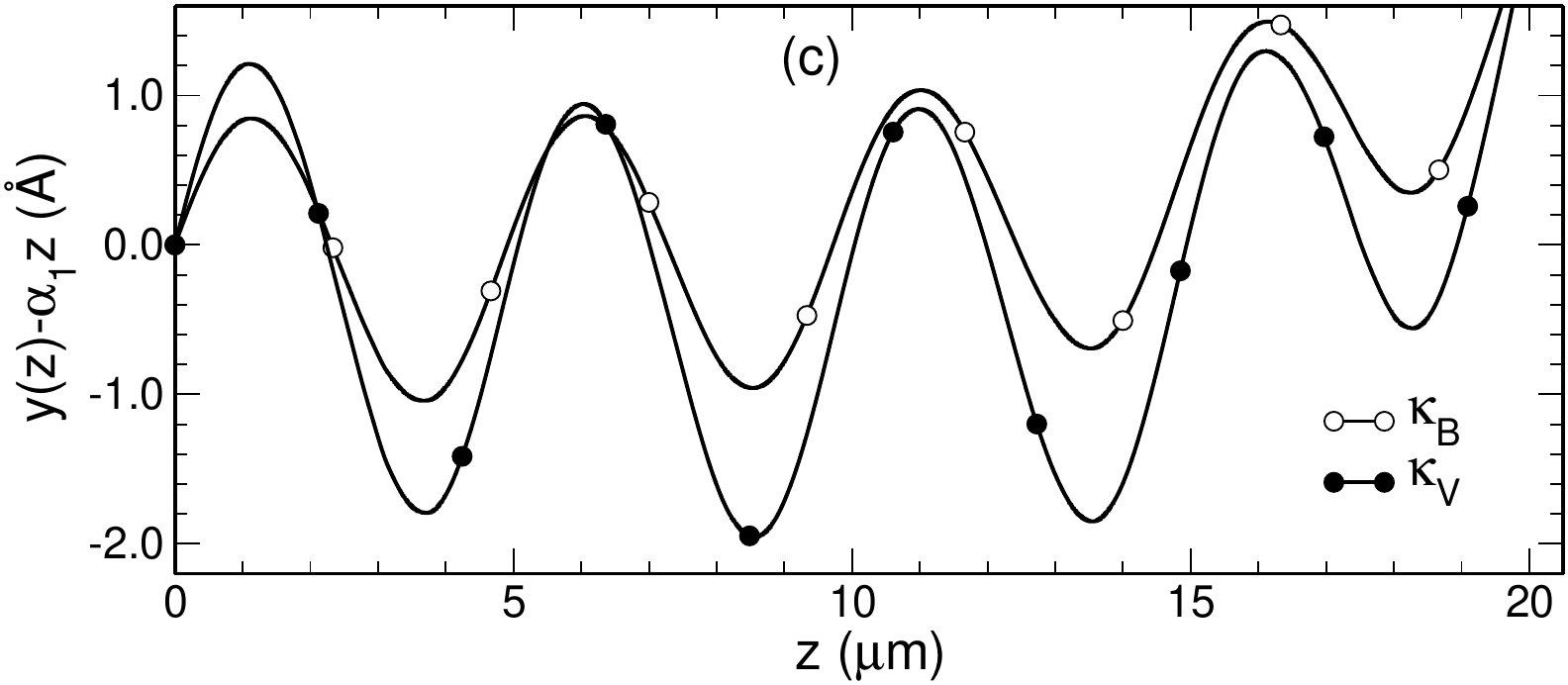}
\vspace*{-1.cm}
\caption{\textbf{(a)}
Boron concentration $n_{\rm B}(Z)$ measured experimentally
(solid line)
and its harmonic approximation used in Ref.
\cite{BackeEtAl:arXiv_2504.17988} (dashed line).
\textbf{(b)}
 The profiles $y(z)$ shown as solid lines,
are calculated using Eq. (\ref{eq:profile})
with the experimental boron concentration and
for two values of the coefficient $\kappa$ mentioned in the text:
$\kappa_{\rm B}$  (B-profile) and $\kappa_{\rm V}$  (V-profile).
The dashed lines show the 'centreline' of the profiles.
Also shown for each profile are the angles $\alpha_1$
and $\alpha_0$ (in the inset), which are the angles
between the $z$-axis and the centreline
and the tangent line at $z = 0$, respectively.
\textbf{(c)} The deviation of the profiles from their
centerlines.
}
\label{Figure02.fig}
\end{figure*}

Eq. (\ref{eq:profile}) facilitates the precise determination of
the bending
profile for any given functional dependence of the dopant concentration,
$n_B(Z)$.
The values of $ n_B(Z) $ employed herein were derived from experimental
measurements.
The dependence $n_B(Z)$ is shown in Fig. \ref{Figure02.fig}(a)
as a solid line.
The minimum and maximum values of the boron concentration are
$n_{\min} = 1 \times 10^{20} \,\mathrm{cm}^{-3}$
and
$n_{\max} = 1 \times 10^{21} \,\mathrm{cm}^{-3}$, respectively.
The dashed line shows the approximation,
$0.5\left[n_{\max}+n_{\min}
- \left(n_{\max}-n_{\min}\right)\cos(2\pi Z/\lambda_{\rm B})\right]$
with $\lambda_{\rm B}=3.54$ $\mu$m, used in Ref.
\cite{BackeEtAl:arXiv_2504.17988} (see Appendix A in
\cite{BackeEtAl:arXiv_2504.17988}).

Figure \ref{Figure02.fig}(b) shows the periodic profiles $y(z)$ (solid lines)
of the (110) planes in the boron-doped layer.
The profiles were calculated for $\kappa_{\rm B}$
\cite{BrazhkinEtAl-PRB_v74_140502_2006} and $\kappa_{\rm V}$
\cite{Vegard:ZPhysik_v5_p17_1921}
using the experimentally measured concentration $n_B(Z)$
\footnote{For brevity, these profiles are hereafter referred to
as 'B-profile' and 'V-profile'.}.
The dashed lines represent their 'centrelines' (the undulator axis)
and the values of $\alpha_{1}$ indicate the angle between the
undulator axis and the $z$ direction.\footnote{We note that the value
of $\alpha_{1}$ for the B-profile differs from
\textcolor{black}{$-174$ $\mu$rad indicated}
in Ref. \cite{BackeEtAl:arXiv_2504.17988} for the ideal
harmonic dependence shown in Fig. \ref{Figure02.fig}(a) and for
$\kappa \approx \kappa_{\rm B}$.}
At $z = 0$, the angles $\alpha_{0}$ between the tangents to the profiles
and the $z$-axis are indicated in the inset.
This angle defines the optimal entrance direction into the crystal channels.

Figure \ref{Figure02.fig}(c) shows the variation of each profile
with respect to its centerline.
They are calculated by subtracting the linear dependencies
$\alpha_1z$ from $y(z)$.


Building on a previous line of research
\cite{UHA-MBN:NIMB_v556_165515_2024,UHA-MBN:JPB_v57_175203_2024},
a rigorous study was conducted by generating a large number
$N_0$ ($\sim 10^4$) of
statistically independent trajectories for the projectile particles
(in this case, electrons).
The motion of an ultra-relativistic charged particle  within an atomic
environment is modeled by numerically integrating the relativistic
equations of motion taking into account the interaction with the
electrostatic field generated by individual atoms
\cite{MBN_ChannelingPaper_2013}.
The atomic potentials are computed within the framework of the
approximation due to Moli\`{e}re \cite{Moliere}.
The general methodology implemented in \MBNExplorer for generating
particle trajectories in a crystalline environment incorporates
(i) randomness in sampling of incoming particles
with respect to the transverse coordinates and velocities
at the crystal entrance (due to the crystal orientation and
the beam emittance),
(ii) deviations of the lattice
atoms from their nodal positions caused by thermal vibrations, and
(iii) a random change in the direction of the particle’s velocity
in the events of inelastic scattering from the crystal atoms
\cite{SushkoKorolSolovyov:NIMB_v569_165911_2025}.
As a result, each trajectory corresponds to a unique crystalline
environment so that all simulated trajectories are statistically
independent and can be analysed further to quantify the channelling
process and characterise the emitted radiation
(see a review article \cite{KorolSushkoSolovyov:EPJD_v75_p107_2021}
where this methodology is applied to a number of case studies).
%

\begin{figure*}
\centering
\includegraphics[width=1.0\textwidth]{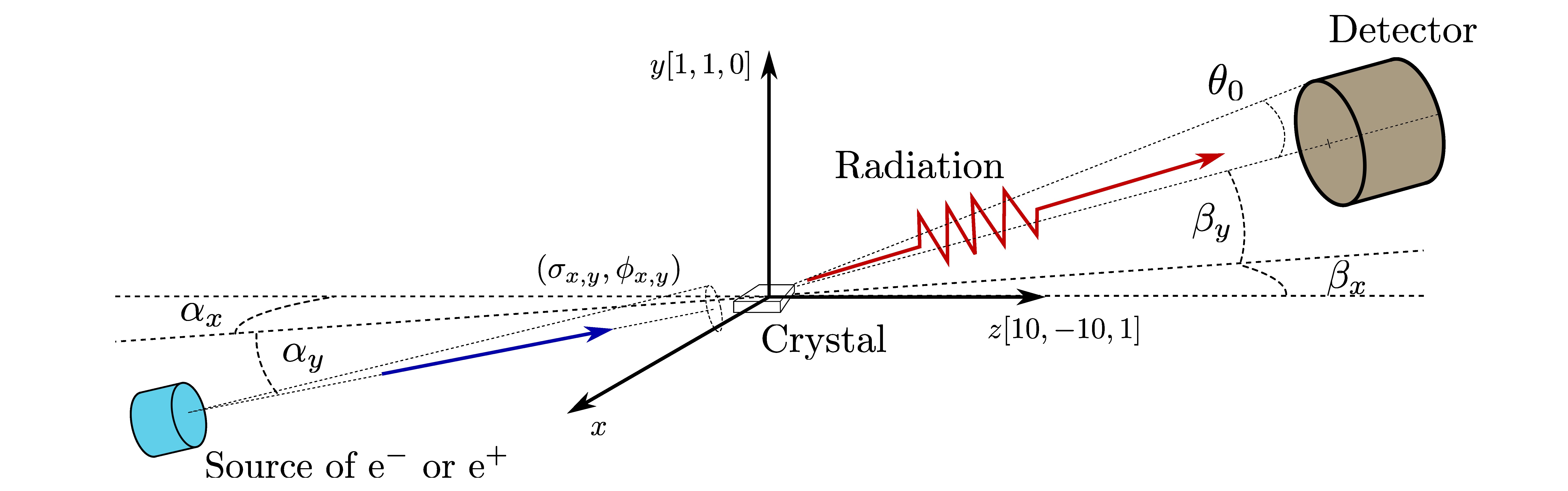}
\caption{Schematic representation of the system in study.
In reference frame $(x,y,z)$, the $y$ and $z$ axes are aligned with the
[110] and [10,-10,1] crystallographic directions in the substrate,
respectively.
The direction of the incident beam angles is characterised by small
angles $\alpha_x$ and $\alpha_y$ measured with respect to the $z$ axis.
The direction of the radiation emission is described by the angles
$\beta_x$ and $\beta_y$;
$\theta_0$ stands for the \textcolor{black}{aperture} cone.
The beam sizes and divergences in the $(xy)$ plane are notated as
$\sigma_{x,y}$ and $\phi_{x,y}$, respectively.
\textcolor{black}{We note that the changes in their values due to
finite but small values of the angles $\alpha_{x,y}$ are neglected.}
}
\label{Figure01.fig}
\end{figure*}

The various parameters, used in the simulations, are indicated in
Fig.  \ref{Figure01.fig}, which shows a schematic representation of the
system under study.
The angles $\alpha_j$ (here and below $j=x,y$) correspond to the direction
of the incident beam.
The parameters $(\sigma_j, \phi_j)$ describe the beam sizes and divergences.
The angles $\beta_j$ indicate the direction along which
the radiation is detected.
The angle $\theta_0$ denotes the reception aperture angle.

Based on the analysis and results of the calculations presented in
\cite{KorolSolovyov:NIMA_v1083_171160_2026},
one can expect the optimal parameters that maximise the CUR intensity
to be as follows: $\alpha_x=\beta_x=0$, $\alpha_y=\alpha_0$, and
$\beta_y \approx \alpha_1$,
\textcolor{black}{see Fig. \ref{Figure02.fig}(b)}.
In the latter relation, the approximate equality is due to uncertainty
in defining the centreline of the profile for a non-harmonic dependence,
$n_B(Z)$.
For ideal periodic bending, the approximate equality sign can be
replaced with an equal sign.

\section{Results and discussion \label{Results}}

\textcolor{black}{In the simulations,
the electron beam was incident on the boron-doped layer
of the hetero-crystal.
This corresponds to the experimental setup \cite{BackeEtAl:arXiv_2504.17988}.
}
%
For each profile shown in Fig. \ref{Figure02.fig}(b), simulations of electron
passage were carried out together with calculations of the emitted radiation
spectra.
The trajectories were simulated across the entire hetero-crystal,
including the thick substrate, for different beam–crystal alignments,
i.e. varying the angles $\alpha_{x,y}$, see  Fig. \ref{Figure01.fig}.
The photon emission spectra were calculated for
\textcolor{black}{the aperture $2\theta_0=268$ $\mu$rad
\cite{BackeEtAl:arXiv_2504.17988} and for}
several sets of angles
$\beta_{x,y}$.\footnote{To avoid confusion, we note that the $x$ and
$y$ directions chosen in this paper correspond to the $y$ and $x$
directions adopted in Ref. \cite{BackeEtAl:arXiv_2504.17988}.}

For each simulated trajectory, the spectral-angular distribution of
radiation $\d^3 E_n/\d(\hbar \omega) \d\Omega$ ($n=1,\dots,N_0$)
emitted within the solid
angle $\d\Omega = \theta \d\theta \d\phi$
(where $\theta\ll 1$ and $\phi$ are the polar angles of the emission
direction)
is calculated numerically following the algorithm outlined in
Refs. \cite{ChannelingBook2014,MBN_ChannelingPaper_2013}.
The spectral distribution $I$ of energy radiated
along any chosen direction
within the cone $\theta_0$ and averaged over all trajectories is
calculated as follows:
\begin{equation}
I(\beta_x,\beta_y)
\equiv
{\d E(\theta\le\theta_0) \over\d(\hbar \omega)}
=
{1 \over N_0}\sum_{n=1}^{N_0}
\int\limits_{0}^{2\pi}\d\phi
\int\limits_{0}^{\theta_0}\theta \d\theta
{\d^3 E_n \over \d(\hbar \omega)\d\Omega}\,.
\label{eq:dE}
\end{equation}
Here the angles $\beta_x$ and $\beta_y$ characterized the direction of
the emission.

Let us introduce a notation for the
difference between spectral distributions calculated for two sets of the
angles $\beta_{x,y}$:
\begin{equation}
\Delta I(\beta_{x2},\beta_{y2}; \beta_{x1},\beta_{y1})
=
I(\beta_{x2},\beta_{y2})
-
I(\beta_{x1},\beta_{y1})\,.
\label{eq:DeltaE}
\end{equation}

\subsection{Simulated trajectories \label{Trajectories}}

\begin{figure*}
\centering
\includegraphics[clip,width=0.49\textwidth]{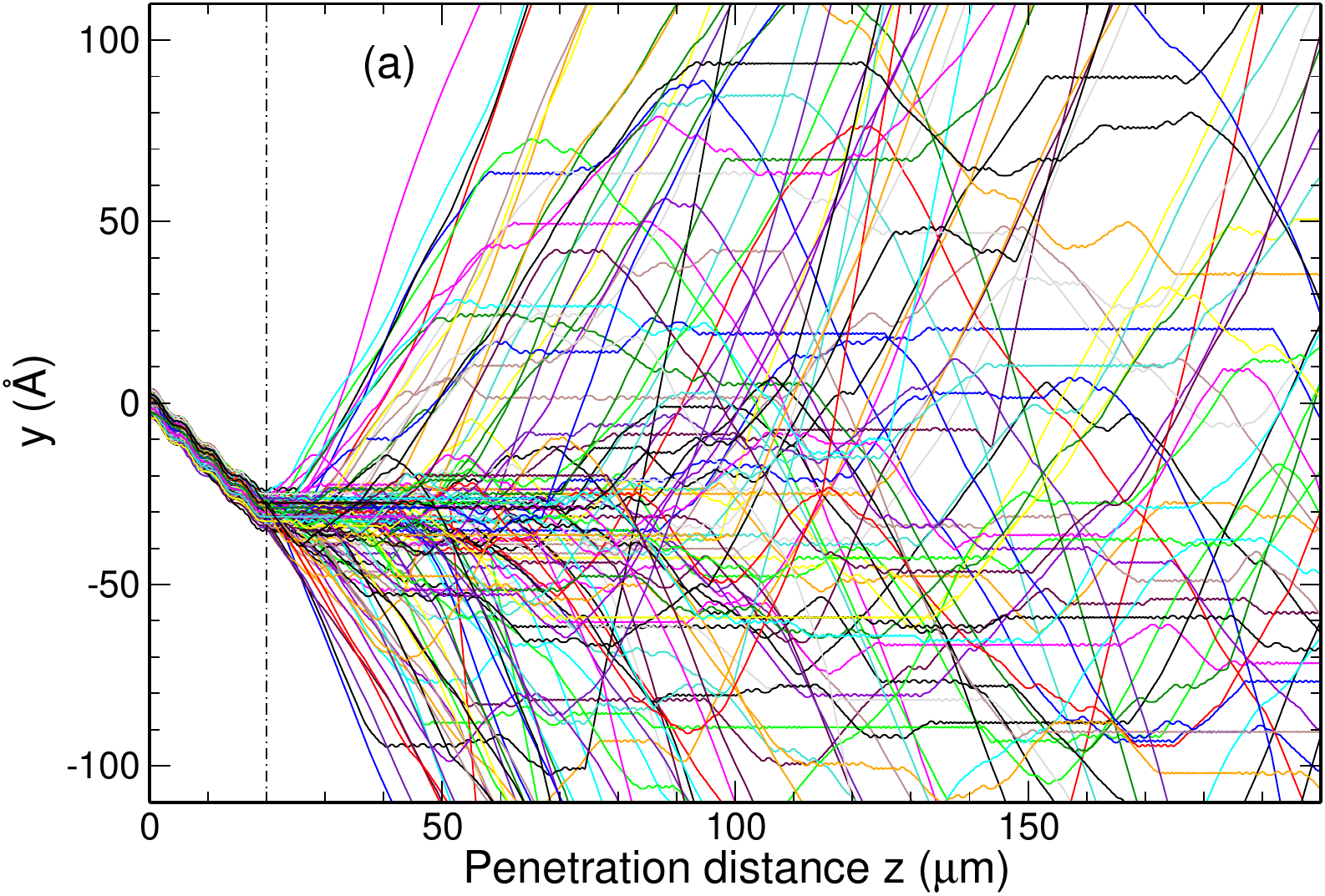}
\includegraphics[clip,width=0.475\textwidth]{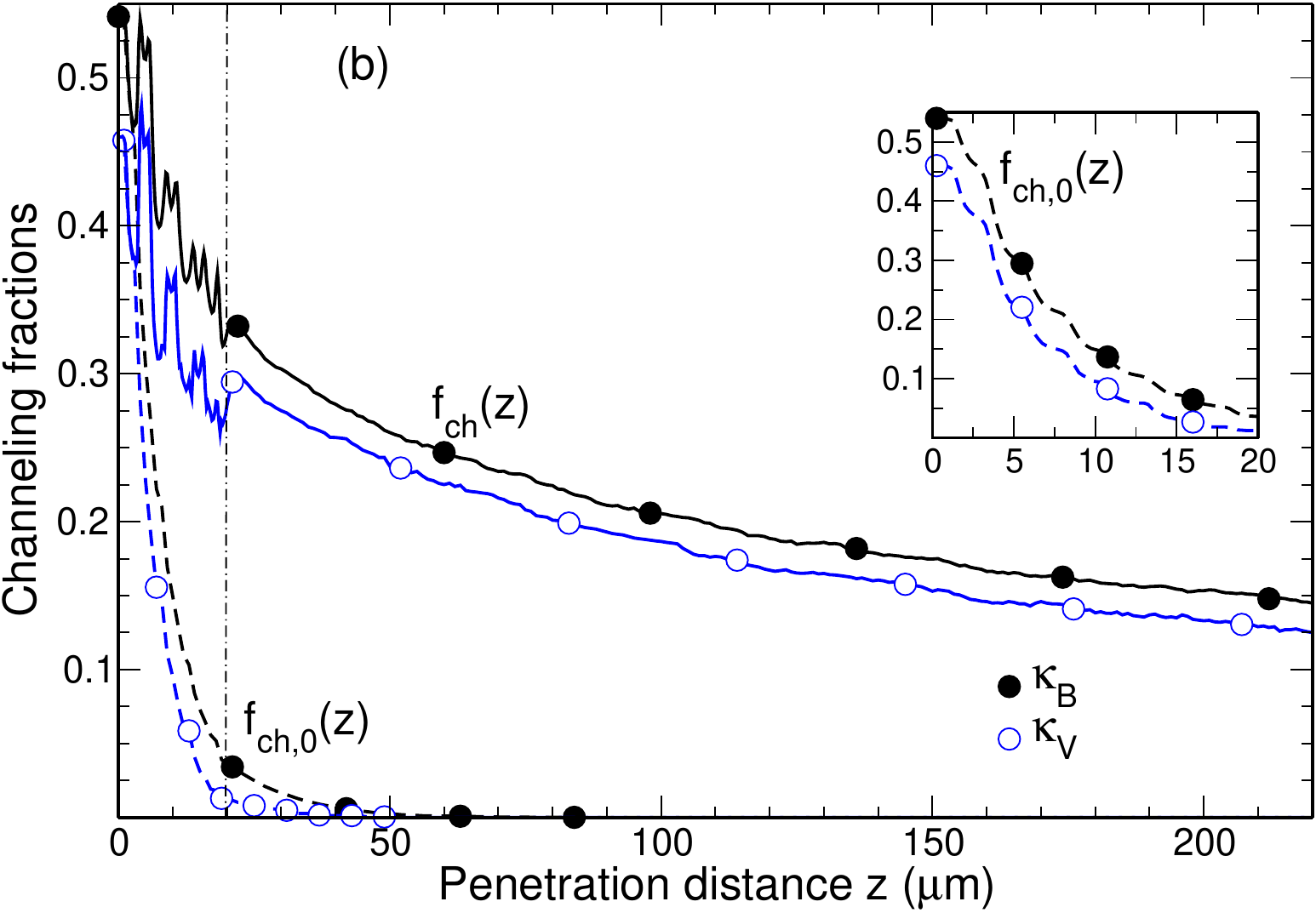}
\caption{\textbf{(a)} Several illustrative trajectories of
electrons channelling
through the entire boron-doped layer (to the left of the vertical dashed
line) and continuing to propagate further into the substrate.
 \textbf{(b)} Channeling fractions $f_{\rm ch, 0}(z)$ (dashed lines)
and $f_{\rm ch}(z)$ (solid lines)
calculated for the B-profile ($\kappa_{\rm B}$)
and
and V-profile ($\kappa_{\rm B}$)
marked with closed and open circles, respectively.
In both panels, the vertical dash-dotted line at $z = 20$ $\mu$m marks
the interface between the doped layer and the substrate.
}
\label{Figure03.fig}
\end{figure*}

Figure \ref{Figure03.fig}(a) shows several selected trajectories of
electrons that were accepted into the channelling mode at the crystal
entrance and travelled through the entire boron-doped layer (penetration
distances to the left of the vertical dash-dotted line at $z = 20$ $\mu$m).
The trajectories correspond to the B-profile and for the
beam-crystal geometry with $\alpha_x=0$, $\alpha_y=\alpha_0$
(see Fig. \ref{Figure01.fig}).%
The figure also highlights the key features of particle propagation
in the substrate ($z > 20$ $\mu$m),  including channelling
and over-barrier motion,
as well as dechannelling and rechannelling events.

To quantify the impact of the dechannelling and rechannelling effects,
one can calculate the channelling fractions
$f_{\rm ch,0}(z) = N_{\rm ch,0}(z)/N_0$
and $f_{\rm ch}(z) = N_{\rm ch}(z)/N_0$
\textcolor{black}{\cite{KorolSushkoSolovyov:EPJD_v75_p107_2021}}.
Here $N_0$ is the total number of incident particles,
$N_{\rm ch,0}(z)$ is the number of the accepted particles that channel
from the crystal entrance up to a distance $z$ in the bulk, and
$N_{\rm ch} (z)$ is the total number of particles moving
in the channelling mode at $z$.
The value of $f_{\rm ch,0}(0)$ determines the acceptance rate,
representing the fraction of the particles accepted at the entrance
out of the total number of particles.

Figure \ref{Figure03.fig}(b) shows the dependence of the channelling
fractions
$f_{\rm ch,0}(z)$ (dashed curves) and $f_{\rm ch}(z)$ (solid curves)
calculated for the B-profile ($\kappa_{\rm B}$) and
and V-profile ($\kappa_{\rm B}$).

On average, the bending curvature of the V-profile is higher, resulting
in a lower acceptance rate ($f_{\rm ch,0}(0)=0.46$ versus 0.54 for the
B-profile) and a steeper decrease in both fractions.
The fraction of accepted particles decrease monotonically, and its
behaviour within the boron-doped layer is shown in the inset.
For both profiles, the number of particles accepted and channelled through
the entire 20-microns-thick layer is at the level of a few percent.
Upon entering the substrate (to the right of the vertical dash-dotted
line),
all of these particles dechannel within a
100 $\mu$m segment.
The values of $f_{\rm ch}(z)$ are notably higher than those of
$f_{\rm ch,0}(z)$ due to the rechannelling.
In the boron-doped layer, the oscillations in $f_{\rm ch}(z)$,
characterised by periodic decreases and recoveries, reflect
alternating dechannelling and rechannelling events.
The minima in $f_{\rm ch}(z)$ occur in the vicinity of the bending extrema
where the curvature is greatest.
The subsequent partial recovery of $f_{\rm ch}(z)$ corresponds to
rechannelling, which predominantly occurs near the midpoints of the bending
profile, where the curvature is minimised.

The distributions $f_{\rm ch,0}(z)$ and $f_{\rm ch}(z)$ permit the calculation
of two quantities that characterise the distances over which particles move in
the channelling mode through a crystal of thickness $L$.
One of these, denoted as $L_{\rm p}$, represents the average length of the
initial channelling segment for accepted particles:
$L_{\rm p} = N_{\rm acc}^{-1}\int_0^L N_{\rm ch,0}(z) \d z
= f_{\rm ch,0}^{-1}(0)\int_0^L f_{\rm ch,0}(z)(z) \d z$,
where $N_{\rm acc}=f_{\rm ch,0}N_0$ is the number of such particles.
The other, $L_{\rm tot}$, is the average of the sum of all
channelling segments per trajectory:
$L_{\rm tot} = \int_0^L f_{\rm ch}(z) \d z $.
These quantities calculated for the B-profile are
$L_{\rm p}=8.8$ $\mu$m and $L_{\rm tot}=59.6$ $\mu$m.
The values for the V-profile are 5.9 and 55.1 microns, respectively.

The channelling radiation is emitted by a particle which moves in the
channelling mode in any part of the hetero-crystal.
\textcolor{black}{For given aperture and divergence,} the intensity of the
channelling radiation (per trajectory) is proportional to  $L_{\rm tot}$.
On the other hand, CUR is predominantly emitted by the accepted
particles that channel in the boron-doped layer so that its intensity
scales with $L_{\rm p}$.
The strong inequality \textcolor{black}{$L_{\rm tot} \gg L_{\rm p}$} holds
for both profiles.
Therefore, one can expect the peak intensity of the channelling radiation to
be much higher than that of CUR.

\subsection{Spectral distribution of radiation \label{Spectra}}

In this section, we compare the calculated spectral distributions
with the experimentally measured dependencies \cite{BackeEtAl:arXiv_2504.17988}.
The emission cone $\theta_0=133.8$ $\mu$rad was chosen to perform
the  integration over $\theta$ in Eq. (\ref{eq:dE}).
This value corresponds to the aperture
$2\theta_0 = 2\, \mbox{mm}/7473\, \mbox{mm}$
calculated from the data presented in Ref.
\cite{BackeEtAl:arXiv_2504.17988}.

In the opening paragraphs of Sect. III of Ref.
\cite{BackeEtAl:arXiv_2504.17988}, the optimal direction for detecting
CUR is discussed in connection with the
bending profile obtained for the harmonic variation of the boron
concentration (see the dashed curve in Fig.  \ref{Figure02.fig}(b)).
In particular, the angles $\beta_x=0$ and $\beta_y=-174$ $\mu$rad
were identified.
The latter value corresponds to the direction of the axis of the ideal
periodic bending profile calculated with the
$\kappa=\kappa_{\rm B}$ coefficient
from Ref. \cite{BrazhkinEtAl-PRB_v74_140502_2006}.
The experimentally measured spectrum clearly showing the presence of
the CUR peak refers to a different angle: $\beta_y=-200$ $\mu$rad
(see the green curve in Fig. 4(a) in Ref \cite{BackeEtAl:arXiv_2504.17988}).
In Ref. \cite{BackeEtAl:arXiv_2504.17988} this spectrum is compared
to that measured for $\beta_y=200$ $\mu$rad,
which is far from the bending profile axis, and thus is mainly due to
the channelling radiation (ChR) emitted in the substrate.
In both cases, the value of the angle $\beta_x$, which was actually
present in the experiment, is not specified.
In the experiment, the crystal was mounted on a goniometer which allowed
for its rotation in all "\dots spatial directions with an accuracy of
35 $\mu$rad" (see the description of the experimental setup in
Sect. IIB in Ref. \cite{BackeEtAl:arXiv_2504.17988}).
Thus the quoted values of the angles
should be considered with the uncertainties,
which should be at least equal to the indicated accuracy.

The experimental spectra refer to a number of events (detected photons)
measured in sr$^{-1}$MeV$^{-1}$.
To ensure an accurate comparison of the
the present results and the experimental results \cite{BackeEtAl:arXiv_2504.17988}
the  following procedure was adopted:
(i) the experimental spectra, digitised from Figs. 4(a) and (b)
of Ref.\cite{BackeEtAl:arXiv_2504.17988},
were multiplied by the photon energy measured in MeV;
(ii) the resulting dependencies and the calculated distributions
were normalised to have the same area within the the photon energy
interval of $0-20$ MeV.

\textcolor{black}{
The spectral distributions, calculated for both B- and V-profiles, were
compared to the experimentally measured ones.
This comparison reveals that the use of the B-profile provides more
consistent agreement with the experimental observations.
Therefore, in the rest of this paper we present the results that
refer to
the the B-profile only.
}

\begin{figure*}
\centering
\includegraphics[clip,width=0.495\textwidth]{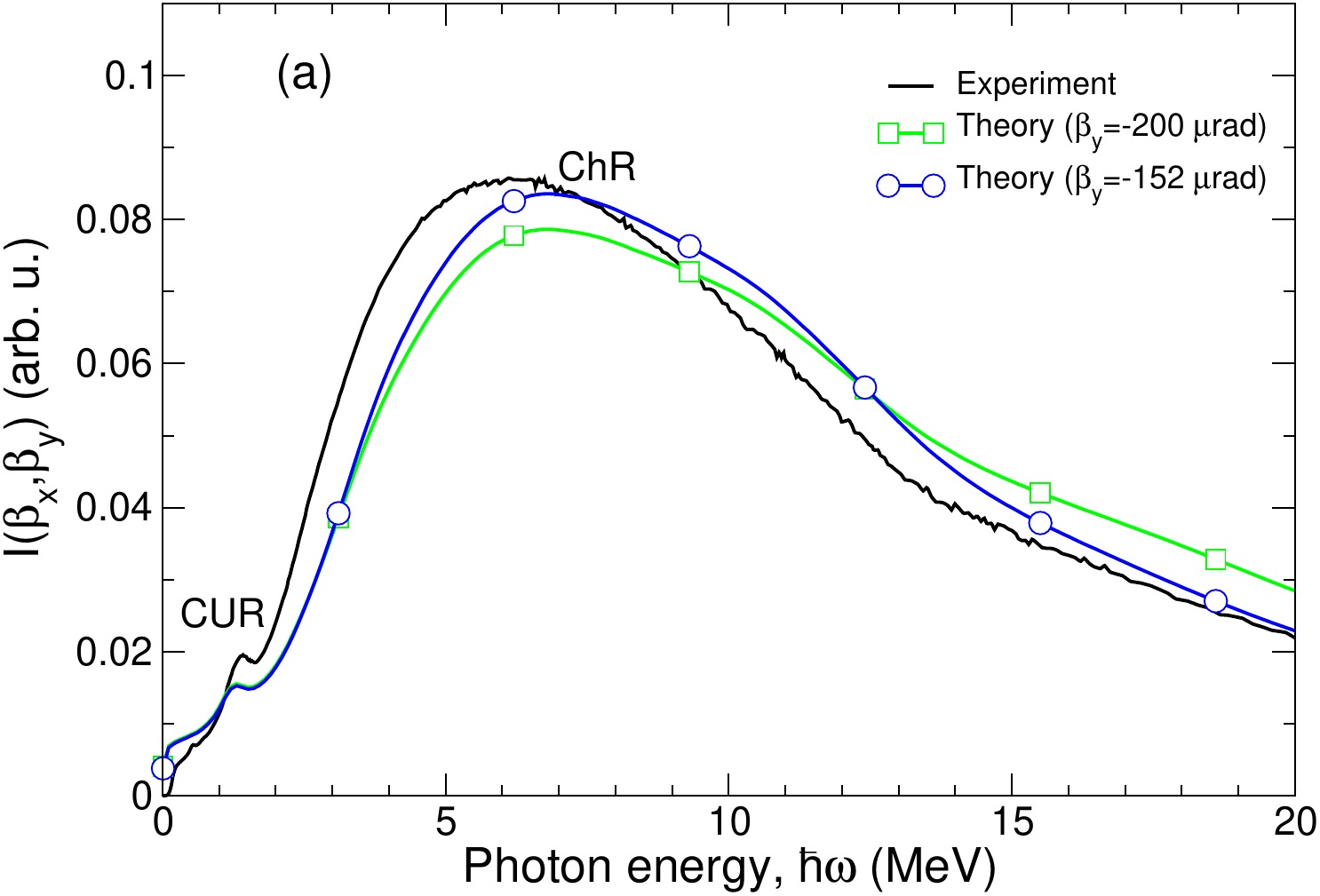}
\includegraphics[clip,width=0.495\textwidth]{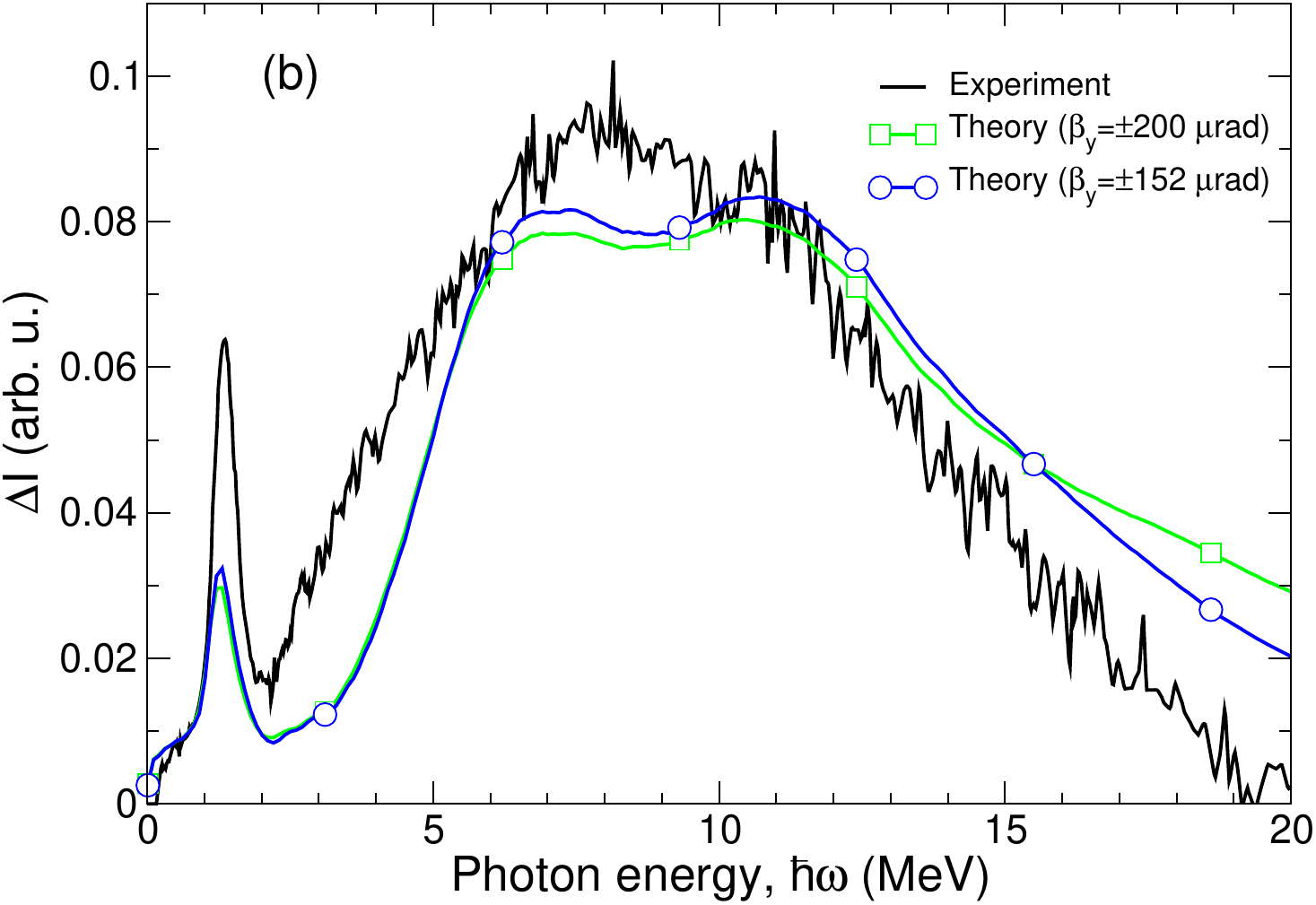}
\caption{
\textbf{Upper row:}
\textbf{(a)} The experimental \cite{BackeEtAl:arXiv_2504.17988}
and calculated spectral distributions
$I(\beta_x,\beta_y)$ of radiation per an 855 MeV
electron in the hetero-crystal.
The experimental dependence corresponds to $\beta_y=-200$ $\mu$rad with
$\beta_x$ not specified.
The calculated dependence corresponds to the angles
$(\beta_x,\beta_y)=(0,-200)$ and  $(0,-152)$ $\mu$rad drawn
in the coloured curves with open circles and squares, respecttively.
\textbf{(b)} The difference $\Delta I =I(0,\beta_y) - I(0,-\beta_y)$ which
emphasises the enhancement of radiation in the low-energy part of the
spectrum due to CUR.
In both panels, all curves are normalised to the unit area.
}
\label{Figure04.fig}
\end{figure*}

Figure \ref{Figure04.fig}(a) compares the experimental spectrum measured
at $\beta_y=-200$ $\mu$rad with the calculated spectrum
$I(\beta_x,\beta_y)$ with $(\beta_x,\beta_y)=(0,-200)$ $\mu$rad (green
curve with open circles) and $(0,-152)$ $\mu$rad (blue curve with open
squares).
The value of $\beta_y=-152$ $\mu$rad corresponds to the direction of
the profile's centerline (see Fig. \ref{Figure02.fig}(b)).
Other parameters used in the simulations include
\textcolor{black}{(i) $(\alpha_x,\alpha_y)=(0,-33)$ $\mu$rad,
which correspond to the
incident beam aligned with the profile's tangent at the entrance,
see Fig. \ref{Figure02.fig}(b)};
and (ii) the beam divergence $(\phi_x,\phi_y) = (20,10)$ $\mu$rad.
The CUR peak appears in the low-energy part of the spectrum.
In the calculated \textcolor{black}{dependencies} it is located at 1.31 MeV
being red-shifted from 1.42 MeV, as in the experimental
spectrum.\footnote{Note that the latter value is larger than
the 1.30 MeV value identified
in Ref. \cite{BackeEtAl:arXiv_2504.17988} for the peak in the
spectral distribution of the emitted photons, $\d N/\d(\hbar\om)$.
The shift in the peak's position is due to the photon energy, which acts
as a factor relating the number of photons to the radiated energy, $I$,
as considered in the present paper:
$\d N/\d(\hbar\om) = (\hbar\om)^{-1} I$.}
For the shown interval of photon energies, the spectrum
is dominated by the broad peak of ChR.
It can be seen that the maximum of \textcolor{black}{ChR in}
the calculated spectrum
is blue-shifted with respect to the experimental one.
The overall agreement with the experimental data is better for
the dependence calculated for $\beta_y=-152$ $\mu$rad.

To emphasise the enhancement of the radiation in the low-energy part of
the spectrum due to CUR, we adopted the methodology outlined in
Ref. \cite{BackeEtAl:arXiv_2504.17988}.
This involves subtracting the background radiation collected within
the same aperture, but in the direction
$(\beta_x,\beta_y)=(0,152)$ $\mu$rad.
The result is shown in \ref{Figure04.fig}(b) together with the
corresponding experimental result.
Both dependencies are normalised to the unit area.
In addition to the shifts noted above, it is seen that the calculated
peak value of CUR is approximately two times lower than the experimental
value.

\begin{figure}
\centering
\includegraphics[clip,width=0.75\textwidth]{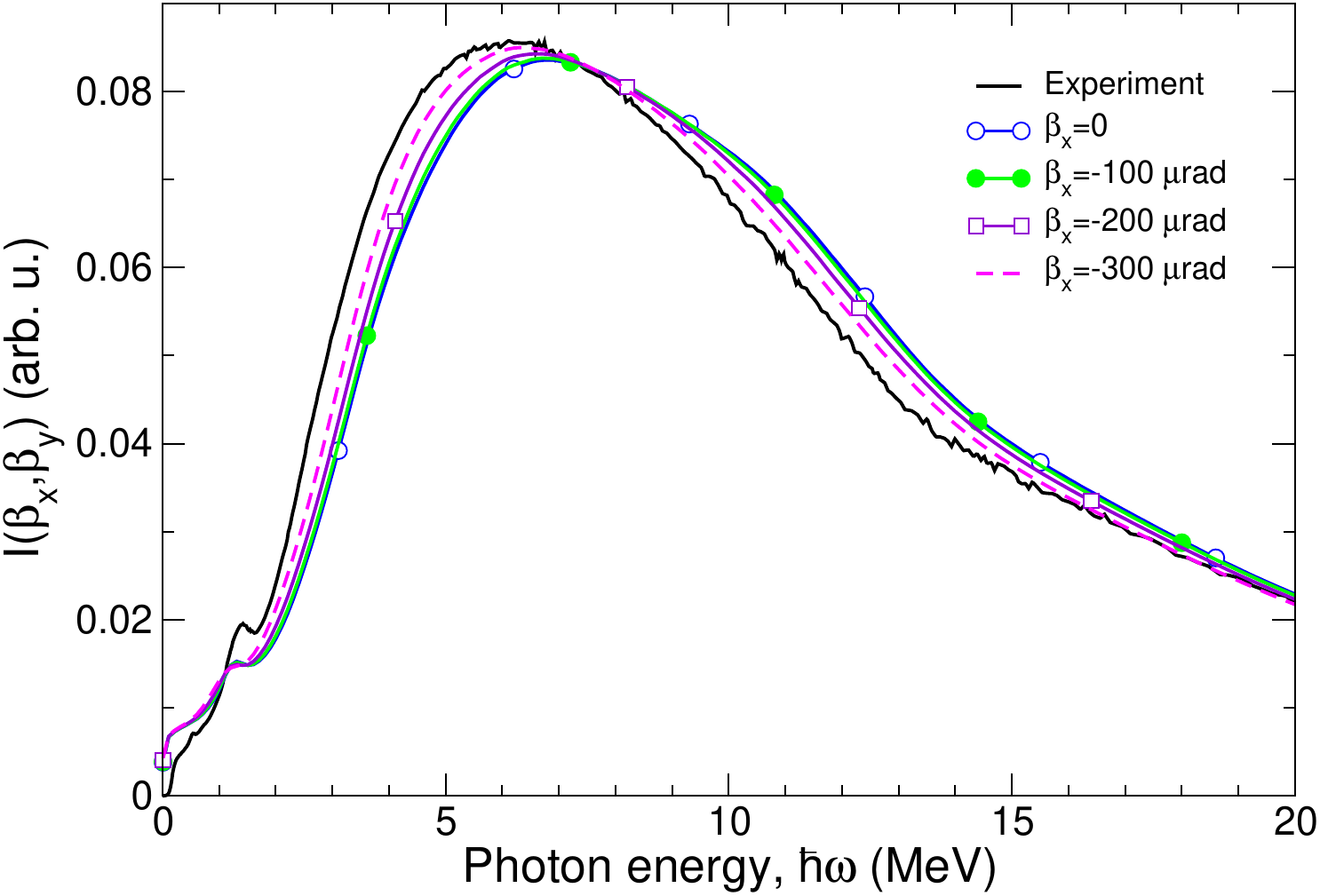}
\caption{
The experimental spectrum measured at
$(\beta_x, \beta_y)=(0,-200)$ $\mu$rad versus
and the spectral distributions $I(\beta_x,\beta_y)$ calculated
for \textcolor{black}{$\beta_x= 0, -100, -200$ and -300} $\mu$rad
and $\beta_y = -152$ $\mu$rad.
All curves are normalised to the unit area.
}
\label{Figure05.fig}
\end{figure}

To investigate the influence of the detector angle $\beta_x$ on the
spectrum, the spectral distributions (\ref{eq:dE}) were calculated for
\textcolor{black}{$\beta_x = 0, -100,  -200$ and -300} $\mu$rad
with $\beta_y$ fixed at -152 $\mu$rad.
An increase in $\beta_x$ shifts the spectrum to
the left, bringing the position of the ChR peak closer to
the experimental result.
This behaviour is clearly illustrated in Figure \ref{Figure05.fig}.
%
%
It is worth noting, given that the involved angles are very small,
that it is challenging from an experimental perspective to position the
detector exactly at $\beta_x = 0$ or at an angle very close to this value.
This may partially explain the observed differences.

Assuming a possible uncertainty in the positioning of the detector,
a series of simulations were performed to establish the
'CUR direction' $(\beta_{x2},\beta_y)$ with $\beta_y=-152$ $\mu$rad
and the 'background direction' $(\beta_{x1},-\beta_y)$,
which provide the best agreement between the calculated
difference of the spectra and the experimentally measured difference
in the vicinity of the CUR peak.

\begin{figure}
\centering
\includegraphics[clip,width=0.75\textwidth]{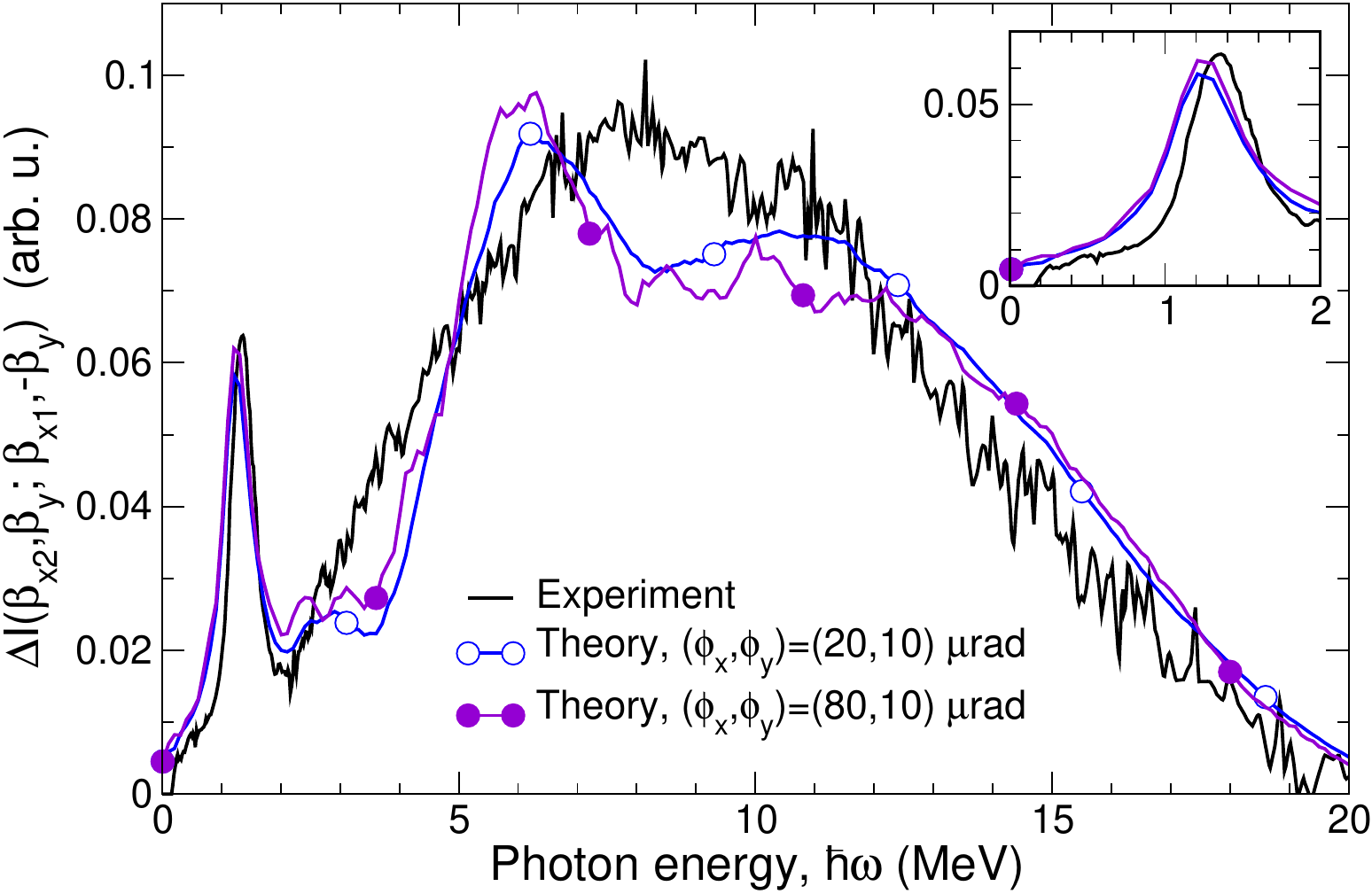}
\caption{
Experiment: the difference between the spectra
measured in the
'CUR direction' with $(\beta_x, \beta_y)=(0,-200)$ $\mu$rad and
in the 'background direction'  (0,200) $\mu$rad
\cite{BackeEtAl:arXiv_2504.17988}.
Theory: the difference between the spectra in the
'CUR direction' (\textcolor{black}{-200},-152) $\mu$rad and in the
'background direction'  with (0,152) $\mu$rad
calculated for two sets of the
beam divergence $(\phi_x,\phi_y)$ as indicated in the legend.
The inset shows the dependencies in the vicinity of the CUR peak.
All curves are normalised to the unit area.
}
\label{Figure067.fig}
\end{figure}

The result  is shown in Fig. \ref{Figure067.fig}, which
compares the experimentally determined difference with that
calculated for (\textcolor{black}{-200},-152) $\mu$rad
(the 'CUR direction') and
(0,152) $\mu$rad (the 'background direction') using Eq. (\ref{eq:DeltaE}).
The calculations were performed for two sets of the divergence
$(\phi_x,\phi_y)$, which are indicated in the legend.
We note that, as the divergence is not specified in Ref.
\cite{BackeEtAl:arXiv_2504.17988}, it can be considered as
another tunable parameter for comparing experimental and
theoretical results.
Overall, both sets of the divergence values reproduce the
experimental trend with comparable accuracy,
although in the CUR region (see the inset) the agreement is
slightly better for $\phi_x = 80$ $\mu$rad.

\section{Conclusions \label{Conclusion}}

In this study, we have presented the results of numerical simulations of the
channeling and photon emission processes for an 855 MeV electron beam incident on
a diamond hetero-crystal consisting of two segments:
(i) a (quasi-)periodically boron-doped diamond layer of thickness
$L\approx 14$ $\mu$m  placed atop of
(ii) a single-crystal diamond substrate of $L=194$  $\mu$m.
The crystalline target parameters (substrate and layer thickness,
number of periods in the layer, and doping profile) and its
orientation used in the simulations
correspond to the experimental conditions cited in Ref.
\cite{BackeEtAl:arXiv_2504.17988}.

Periodic doping along the [100] direction results in periodic bending
of the (110) planes.
The bending profile was calculated using the methodology outlined
previously \cite{KorolSolovyov:NIMA_v1083_171160_2026} and the
experimentally measured boron concentration data.
These calculations, performed for the two values of the coefficient
$\kappa$ available in the literature, enabled us to determine the
precise orientation of the undulator axis relative to the
crystallographic directions in the substrate.
This information is essential for determining the optimal direction of
the incident beam and the position of the detector that collects the
emitted radiation.

The electrons incident on the layer along the (110) planes and accepted
in the channeling mode emit CUR, which is most intense within the cone
$\theta_0\approx 1/\gamma$ along the undulator axis.
To maximise the number of accepted particles the beam must be aligned with the
tangent to the bent (110) plane at the entrance.
Statistical analysis of the trajectories has shown that vast majority of the
accepted particles dechannel within the doped layer and enter the substrate,
moving in the overbarrier mode.
However, frequent rechanneling events result in the total
length of the channeling segments (per trajectory) being much larger than
in the  layer.
Consequently, the channeling radiation predominantly emitted from the
the substrate is more intense than the CUR.

In this paper, we analyse the dependence of the spectra of emitted radiation on a
number of angular variables that characterise the beam-crystal and
crystal-detector geometries.
Some of these angles were specified in the experimental setup description,
while others were either not quoted explicitly or quoted without indicating
the uncertainties.
The results show that factors like the orientation of the
incident beam, and the direction in which the radiation is detected, significantly
affect the radiation intensity.
The similarities and discrepancies between experimental and numerical simulation
spectroscopic results are discussed.
The dependence of the spectrum on the beam divergence $\phi_{x,y}$ was
also investigated.
The best overall agreement with experiment corresponds to
$(\phi_x,\phi_y) = (20, 10)$  $\mu$rad, although in the CUR
region the best agreement is seen for $(\phi_x,\phi_y) = (80, 10)$  $\mu$rad.

%
\section{Acknowledgements}

We acknowledge support by the European Commission
through the N-LIGHT Project within
the H2020-MSCA-RISE-2019 call (GA 872196)
and the EIC Pathfinder Project TECHNO-CLS
(Project No. 101046458).
MMM, GRL, and JRS would like to acknowledge the support of the
national basic and natural sciences program under project code
PN\-223LH010-069.
We also acknowledge the Frankfurt Center for Scientific
Computing (CSC) for providing computer facilities.

\section{Authors contributions}

\textbf{MMM}: Simulations, data processing and analysis.
\textbf{GRL}: Simulations, data processing and analysis.
\textbf{JRS}: Simulations and analysis; Writing -- original manuscript.
\textbf{TNTC}: Data on the doping concentration $n_{\rm B}(Z)$.
\textbf{AVK}: Conceptualization; Methodology, simulations and
analysis; Writing -- review \& editing.
\textbf{AVS}: Project administration; Conceptualization;
Methodology, simulations and analysis; Writing -- review \& editing.

All authors reviewed the final manuscript.

\section*{References}

\end{document}